\documentclass{article}
\usepackage{amssymb}
\usepackage{graphicx}

\newcommand{\ba}{\begin{array}}
\newcommand{\ea}{\end{array}}
\newcommand{\bd}{\begin{displaymath}}
\newcommand{\ed}{\end{displaymath}}
\newcommand{\be}{\begin{equation}}
\newcommand{\ee}{\end{equation}}
\newcommand{\bea}{\begin{eqnarray}}
\newcommand{\eea}{\end{eqnarray}}

\begin{document}
\title{{\begin{flushright}
\end{flushright}}\vskip 0.5 cm {\textbf { Intermediate Scale Inflation
      and Metastable Supersymmetry Breaking}}}
\author{Philippe Brax, Carlos A. Savoy and Arunansu Sil\\
\small{\em Institut de Physique Theorique, CEA, IPhT, CNRS, URA 2306  F-91191
  Gif-sur-Yvette Cedex, France }}
\date{}
\maketitle
\begin{abstract}
We investigate the possibility of obtaining a low scale of
supersymmetry breaking within the ISS framework using a metastable
vacuum. This is achieved by introducing an $R$
symmetry preserving gravitational coupling of the ISS sector
to a relatively low scale inflationary sector. We find the allowed 
range for the supersymmetry breaking scale, $10^{4}$ GeV $
\lesssim \mu \lesssim 10^8$ GeV, which is low enough to be amenable to
gauge supersymmetry breaking mediation. This scenario is based
upon a so-called hilltop inflation phase whose initial condition problem is
also addressed.
\end{abstract}
\maketitle
\vskip 0.7cm


It has been recently realised by ISS \cite{ISS} that supersymmetry
breaking can be achieved in a metastable vacuum which is separated
from the true supersymmetry preserving vacuum by a barrier that
can guarantee a life-time for the false vacuum which exceeds the
age of the universe. One particular advantage of this setting is
that the IR free magnetic description (which is dual to a 
UV free electric theory) is suitable to study  low energy physics.  
This opens up the
possibility of describing supersymmetry breaking at low energy
(the ISS scale $\mu$) compared to the Landau pole of the magnetic
phase. Within the metastable supersymmetry breaking
framework, it has been recently shown in \cite{aru} that $R$
symmetric gravitational couplings between the supersymmetry
breaking sector and  the inflation one would help determining the ISS
scale. It would also provide a natural explanation for why the
universe should end up in  the metastable minimum instead of the
supersymmetric minimum.

The connection between supersymmetry breaking and inflation may
shed some light on our understanding of scales beyond the standard
model of particle physics. Indeed cosmological observations of the
cosmic microwave background anisotropies single out  a very large
scale close to the GUT scale when interpreted  within the
inflationary paradigm. The magnitude of the temperature
fluctuations is given by the height of the Sachs-Wolfe plateau and
corresponds to $\frac{\delta T}{T} \simeq 6.6 \times 10^{-6}$ \cite{cobe, Spergel:2006hy}.
This translates into a constraint on the inflationary potential
$V_I$,
\begin{equation}
\left (\frac{V_I}{\epsilon} \right )^{1/4} \simeq 6.6 \times 10^{16} ~{\rm GeV},
\label{eqvep}
\end{equation}
where $\epsilon$ is the slow roll parameter defined as $\epsilon =
\frac{M^2_P}{2} \left (\frac{V'}{V} \right)^2$. So depending upon
the value of $\epsilon$ at the time of horizon exit, the
inflationary scale ($V_I^{1/4}$) can be estimated. For example, in
supersymmetric hybrid inflation \cite{Dvali:1994ms} ($\epsilon \sim 10^{-8}$), this
scale turns out to be $10^{15-16}$ GeV, ${\it {i.e.}} \sim$ GUT
scale. We have found in \cite{aru} that this corresponds 
to the scale of supersymmetry breaking in the range of gravity mediation. 
For sufficiently low $\epsilon$, this characteristic scale
would be lower. In particular, we will find that for intermediate
values of $V_I^{1/4}\approx 10^{11}$ GeV, the supersymmetry
breaking scale could be as low as $10^4$ GeV. Of course, this is
within the right ball park for gauge mediation of the
supersymmetry breaking to the MSSM. This could well be hint that
supersymmetry breaking effects appear at low energy and could be
observable at the LHC.

In this letter our aim is to find a metastable supersymmetry
breaking at a low scale ($\mu,$ the ISS scale) which is consistent
with gauge mediation of supersymmetry breaking. 
Following the approach in \cite{aru},  we assume the
existence of two sectors, the inflation and the ISS sector, which
communicate with each other only through gravity\footnote{How the
ISS sector interacts with the MSSM, so that the soft
supersymmetry breaking effect can be seen, is beyond the scope of
our present work. For recent works in this direction, see \cite{mediation1, 
mediation2, mediation3}.} 
(respecting $U(1)_R$ symmetry). The inflation
sector consists of superfields $(\chi, ~\bar\chi)$ and $S$, with
$R$ charges, $R$[$\chi \bar\chi$] = 0 and $R$[$S$] = 2. As we are
dealing with gravitational interactions between the two sectors
specified above, it is quite natural to consider  inflation models
in the framework of supergravity. Therefore we must specify the
K\"ahler potential of the inflation sector. We assume that the
K\"ahler potential is invariant under a shift symmetry of the
inflaton chiral multiplet $\chi, \bar\chi$. The choice of this
shift symmetry is mainly motivated by the solution to the $\eta$
problem\footnote{The shift symmetry is also an essential
ingredient of some string inflation models such as the ones based
on the compactification manifold $K_3\times T^2/Z_2$ where the free
motion of branes along the two torus is translated as a shift
symmetry in the K\"ahler potential\cite{kallosh}.}. Thus the inflaton direction 
$\chi = \bar\chi$ does not receive any mass-squared term $\sim O(H_I)$
the presence of which otherwise would spoil the flatness of the potential. 
The K\"ahler
potential is given by
\begin{equation}
K_- = \vert S\vert^2 + \frac{1}{2} \vert \chi -\bar \chi\vert ^2 +
a_1\frac{|S|^4}{4 M^2_P} + a_2\frac{|\chi - \bar\chi|^4}{4M^2_P} +
a_3 |\chi -\bar\chi|^2 \frac{|S|^2}{2M^2_P},
\label{kahler}
\end{equation}
where we keep higher order terms  whose necessity will
be spelt out later. 

It is a  known fact that to generate inflation
we need to break the exact shift symmetry in order to give a
slope to the inflaton potential. This is achieved by introducing
a higher order (${\it {gravitational}}$) term in the superpotential\footnote{A higher order
breaking of the shift symmetry will also be present in the K\"ahler
potential and  will be crucial in getting rid of the initial
condition problem for hilltop inflation.}.
Now, the inflaton field is defined by $\chi=\bar \chi$, while the $\chi= -\bar \chi$
direction corresponds to a massive field which  plays no role in inflation and can
be discarded from the discussion. Hence we keep only the  inflaton field which,
for convenience, we still denote $\chi$, in the inflationary superpotential that
we write in the form \bea
W_{inf} &=& S \left ( k \frac{\chi^{n}}{M_{\rm P}^{n - 2}} - M^2 \right ),
\label{eqngen} \eea where $n > 2$ and a discrete symmetry identically
transforming $\chi$ and $\bar\chi$ guarantees the form of the
 superpotential.

The ISS sector is described by a supersymmetric $SU(N_c)$ gauge
symmetry with $N_f$ flavors of ${\it {massless}}$ quark-antiquark
pairs in the electric theory. Here $\Lambda$ is the
strong-coupling scale of the theory, below which the theory can be
described as the magnetic dual, $SU(N)$ gauge theory, where $N =
N_f - N_c$ with $N_f$ flavors of magnetic quarks, $q^c_i, \tilde
q^i_c$, ($i = 1\dots N_f$ and $c = 1\dots N$) and a $N_f \times N_f$
gauge singlet superfield $\Phi^i_j$ (the meson field $\Phi = Q
\tilde Q / \Lambda$). The magnetic theory is IR free if $N_c + 1
\le N_f \le \frac{3}{2}N_c$ and has a  superpotential given by
\be W = h{\rm Tr}q \Phi \tilde q, \label{eqn2} \ee for massless
quarks, along with the dynamical superpotential \be W_{dyn} =
N\left(h^{N_f} \frac{{\rm {det}} \Phi}{\Lambda^{N_f -
3N}}\right)^{\frac{1}{N}}, \label{eqn3} \ee where $h = O(1)$. The
R-charges are such that 
$\Phi$ has a $R$-symmetry charge  $R[\Phi] = 2$, $R[Q] = R[\tilde Q] = 1$
up to a baryon number  and $R[q, \tilde q] = 0$.

The interaction between these two sectors can be described (in the
magnetic phase)  by 
\be W_{int} = \lambda \frac{\chi^n}{M^{n-1}_P}\Lambda
{\rm {Tr}} \Phi \label{int}
\ee 
which respects $U(1)_R$ symmetry as well as
the discrete symmetry imposed upon the $\chi, \bar\chi$ fields
(this restricts also the form of  the $\chi$-dependent terms in
the superpotential of the inflation sector as we discussed
before.). Once inflation ends, the $\chi$ field gets
a vev and $W_{int}$  induces a mass term for the electric quarks,
\be W_{\rm ISS} = h{\rm Tr}q \Phi \tilde q - \mu^2 {\rm {Tr}}
\Phi, \label{eqn2} \ee which is the same superpotential as
analysed by ISS with   $\mu^2$  defined as
$\frac{\lambda}{k}\frac{\langle \chi^n \rangle}{M^{n-1}_P} \Lambda$. It turns out that for $\mu \ll
\Lambda$, supersymmetry is broken at the metastable minimum,
$\langle \Phi \rangle = 0, ~ \langle q \rangle= \langle \tilde q
\rangle = \mu$.
In our approach,  the scale of supersymmetry breaking can be
written as 
\be \mu^2 = \sqrt{3}\frac{\lambda}{k} H_{I} \Lambda, 
\label{eqmu} 
\ee 
where
$H_I$ is the Hubble scale during inflation ($ H_I^2 =
\frac{V_{I}}{3 M^2_P}$) as the $\chi^n$ term in the
inflationary superpotential  cancels the vacuum energy during
inflation ($M^4$). We have assumed that  gravity respects the
$R$-symmetry as well as the discrete symmetry imposed upon $\chi$ and $\bar\chi$.
Using the
constraint on $\mu$ from the metastability condition \cite{ISS} ($\mu <
\Lambda$), we find that $H_I < \Lambda$. This means that the only
way of  achieving a low scale of supersymmetry breaking is through
low value of $H_I$, i.e. lowering the scale of inflation. With
supersymmetric hybrid inflation model, it is not
possible to lower the scale of inflation very  much \cite{aru}. 
On the contrary  if we adopt a hilltop type of inflation
model  where the inflaton rolls down from a saddle point towards a
minimum, we can achieve a low value of $H_I$. This leads to a low value of $V_I$ which is then
consistent with the COBE data \cite{cobe} as $\epsilon$ turns out to be very
small (see Eq.(\ref{eqvep})). This model has also  the power of
explaining a low value of the spectral index as obtained from  the
WMAP 5 years data \cite{Spergel:2006hy} $n_s \approx 0.96 \pm 0.014$. We find
that a minimalistic choice for  $n$ is 4. This entails that the
discrete symmetry we have discussed before would be a $Z_4$
invariance, under which both $\chi$ (also $\bar\chi$) carry charges $i$
while $S$ has charge 1.

We are now going to discuss the inflationary scenario in more
detail. We start with the superpotential in Eq. (\ref{eqngen}) for
$n =4$.  The supersymmetric vacuum is given
by $S = 0$ and $k \chi^4 = M^2 M^2_P$. The scalar potential in
supergravity using the K\"ahler potential in eq.(\ref{kahler})  along the inflaton direction
reads \begin{equation}
V \simeq M^4 \left[ 1 - a_1 \frac{|S|^2}{M^2_P} - 2k\frac{|\chi|^4}{M^4_X}
+ k^2 \frac{|\chi|^8}{M^8_X}
 \right ],
\label{eqnv}
\end{equation}
where $M^2_X = M M_P$ and we have chosen $a_1 < -$ 1/3, so that
$S$ receives a positive mass square greater than $H^2 \simeq
M^4/{3M^2_P}$ during inflation and therefore rapidly settles to
zero. Such a class of  potentials\cite{shafinew} has been already considered
\cite{newinf} and happens
to be a good approximation to the dynamics of racetrack inflation
in string theory\cite{marieke}.
Inflation takes place when the field starts close to the origin
($\chi \simeq 0$) where the potential is maximal. From there it
rolls down at a slow rate before eventually settling down at the
supersymmetric minimum far away from the origin, $k\langle  \chi^4
\rangle = M^4_X$. The fact that the inflaton starts from a low
value compared to the Planck scale is an initial condition issue
which will be discussed later.

The slow roll parameters are given by (${\rm {for}} ~|\chi| \ll M_X $) 
\bea
\epsilon & = \frac{M^2_{P}}{2} \left ( \frac{V'(\chi)}{V(\chi)} \right )^2 
& \simeq 32 M^2_P \frac{\chi^6}{M^8_X}~, 
\label{ep}\\
\eta & = M^2_P |\frac{V''(\chi)}{V(\chi)}|
& \simeq 24 M^2_P
\frac{|\chi|^2}{M^4_X}~. 
\label{eta}
\eea 
The field value at the end of inflation, $\chi_f$, is given by $|\eta| \simeq
1$, 
\be \chi_{f} \simeq
\frac{1}{2\sqrt{6k}}\frac{M^2_X}{M_P}. 
\label{eqchi} 
\ee 
The
number of e-foldings, $N$, then relates the initial value of the
inflaton field, $\chi_0$ at the time of horizon exit with $\chi_f$
by 
\be N = \frac{1}{M^2_P}\int^{\chi_0}_{\chi_f} \frac{V
d\chi}{V'} \simeq \left ( \frac{1}{\chi^2_0} -
\frac{1}{\chi^2_{f}}\right ) \frac{M^4_X}{16k M^2_P}, 
\ee 
hence
$\chi_0 \simeq \frac{1}{2\sqrt{2k}(3 + 2N)^{1/2}}
\frac{M^2_X}{M_P}$ where we have used eq.(\ref{eqchi}). The
spectral index is given by \be n_s \simeq 1 - 2\eta \simeq 1 -
\frac{6}{3 + 2N}. \ee With $N = 52$, the resulting\footnote{The
number of e-foldings is related with the scale of inflation by
$N \simeq 60 - \log{\left ( \frac{10^{16} {\rm{GeV}}}{V^{1/4}_{I}}\right)}$.} 
spectral index $n_s \simeq 0.945$ which is within 1$\sigma$
of the central value of the spectral index as recently prescribed
by the WMAP result \cite{Spergel:2006hy}.

The inflation scale is determined by the COBE normalisation
\begin{equation}
(\frac{V_I}{\epsilon})^{1/4} \sim 6.6 \times 10^{16} ~{\rm GeV}.
\label{vbyep}
\end{equation}
Using eqs.(\ref{eqnv}) and (\ref{ep}), $\epsilon$ at the time of horizon exit can be expressed as
\begin{equation}
\epsilon \simeq  \frac{k^2}{16 \left ( 3 + 2N \right )^3}
\left ( \frac{M_X}{M_P} \right )^4.
\end{equation}
Therefore using $V_I = M^4$ and $M^2_X = M M_P$, we find from
eq.(\ref{vbyep}) that  the scale of inflation is
\begin{equation}
M \simeq \frac{4.5}{(3+2N)^{3/2}} \times 10^{14} ~{\rm {GeV}} \sim 10^{11}~{\rm {GeV}}.
\label{eqM}
\end{equation}
Notice that the natural scale $\approx 10^{14} ~{\rm GeV}$  is
reduced, thanks to the e-fold factor $(3+2N)^{3/2}\approx 10^3$.
The initial field value $\chi_0$ is required to be $\chi_0\sim
10^{-4}M_X$, this initial condition issue  will be  discussed at
the end of this letter.

Once inflation is over, the coupling in eq.(\ref{int}) implies
that the SUSY breaking scale ($F_\Phi= \mu^2$) is given by
\begin{equation}
\mu^2= \frac{\lambda}{k} \frac{M^4_X \Lambda}{M^3_P}.
\label{mu2}
\end{equation}
In terms of the Hubble rate during inflation, $H_I \simeq \frac{M^2}{\sqrt{3} M_P}$ 
(in the inflationary scenario
considered above, $H_I \sim 10^4$ GeV), this leads to eq.(\ref{eqmu}).
In order to maintain the metastability condition in the ISS
sector, one has to impose a constraint $\mu < \Lambda$, which in
turn sets a lower bound (along with eq.(\ref{mu2})) on the scale
of supersymmetry breaking as 
\be \mu > H_I \simeq 10^4 {\rm
{GeV}}, \label{muupper} 
\ee 
for $\lambda/k \sim 0(1)$. In the
following we will obtain an upper bound while discussing
reheating at the end of inflation. In a similar fashion to
\cite{aru}, $\Phi$ is also stuck  at origin during inflation due
to the presence of a mass term bigger than $H_I$  due to the
supergravity corrections. Notice that when inflation is over, this
point, $\Phi = 0$, becomes a local minimum (this supersymmetry
breaking minimum appears when $\mu$ becomes non-zero as a result
of displacement of $\chi$ from the top of the potential in the
inflation sector) and so the field does not move. This explains
why the universe should prefer the supersymmetry breaking minimum
rather than the supersymmetric one in the ISS sector when one
considers the evolution of the universe.

At the end of inflation, the inflaton field performs damped
oscillations about the supersymmetric minimum of the inflation
sector and decays. The main decay channel follows from
\begin{equation}
V\supset \vert \frac{\partial W}{\partial \Phi}\vert^2 = \vert
h q\tilde q + \lambda \frac{\chi^4}{M^3_P} \Lambda\vert^2.
\end{equation}
This leads to the decay of $\chi$ into magnetic quarks (since we are already in the magnetic phase)
with the decay width
\be
\Gamma \simeq \frac{h^2\lambda^2}{8 \pi k^{3/2}}  \frac{\Lambda^2}{m_{\chi}}
 \left (\frac{M_X}{M_P} \right )^6,
\ee
where $m_{\chi}$ is the mass of the inflaton, $m_{\chi} = \sqrt{2k} M^2/M_P$.
Thus the reheat temperature $T_R \simeq \frac{1}{7}\sqrt {\Gamma
M_P}$ is given by
\begin{equation}
T_R \simeq  \frac{h \lambda \Lambda}{14\sqrt{2\sqrt{2} \pi}}  \frac{1}{k^{7/8}} \left (
\frac{M}{M_P} \right )^{3/4}.
\end{equation}
Imposing that reheating should take place before the electroweak
transition\footnote{We impose this condition keeping in mind that 
the most popular way of generating the baryon asymmetry of the universe via leptogenesis 
will be ineffective if the reheat temperature after inflation is less than $10^2$ GeV. 
This particular decay chanel we consider does not produce MSSM particles and for that 
the inflaton sector should couple with the MSSM sector which we will not discuss here.} 
$T_R \gtrsim 10^2$ GeV leads to a lower bound $\Lambda
\gtrsim 10^8$ GeV where we have used eq.(\ref{eqM}) and $h \sim
\lambda = O(1)$. Since from
the metastability condition we know $\mu < \Lambda$, it results
into an upper bound on the SUSY breaking scale, $\mu < 10^8$ GeV,
obtained for the lowest value of $\Lambda$. Combining it with
eq.(\ref{muupper}), we find that our scenario constrains the scale
of supersymmetry breaking as follows \be 10^4 ~{\rm {GeV}}
\lesssim \mu \lesssim 10^8 ~ {\rm {GeV}}. \ee In this work, we
have not focused on the mediation mechanism, $i.e.$ how the
supersymmetry breaking will be mediated to the MSSM sector. We
keep this for future work  where we will deal with inflation and a
deformed ISS model of supersymmetry breaking in order to include
$R$ symmetry breaking also.

Let us now come back to the initial condition problem mentioned
previously. Indeed we have assumed that $\chi$ is small initially,
$\sim 10^{-4} M_X$. This calls for an explanation. A first
possibility springs from the fact that prior to inflation, the
universe could be radiation dominated and in a high temperature
phase. Here we present a mechanism following \cite{japanese} which
leads to a satisfactory explanation for the initial condition
problem. To address the initial $\chi$ value, we introduce one or 
more superfields $Y_i$ with $R[Y_i] \neq 0$. They may have interactions
with the MSSM (or extended MSSM) superfields. We also postulate a
higher order shift symmetry breaking term in the K\"ahler potential
which is actually a cross term between\footnote{The shift symmetry preserving term, although
present, will not intervene as the inflaton direction is $\chi =
\bar\chi$.} $Y_i$ and $(\chi + \bar\chi)$,
\begin{equation}
K_+=\sum_{i=1}^P  b_i \frac {\vert \chi+ \bar \chi\vert^2}{2
M_{\rm P}^2} \vert Y_i\vert^2.
\end{equation}
Following the approach in \cite{japanese}, the above term leads to
an interaction, the thermal average of which is given by 
\be b_i
\langle \partial_{\mu} Y_i \partial^{\mu} Y_i \rangle
\frac{\chi^2}{M^2_P} \simeq b_i m^2_{Y_i} (T)
\frac{T^2}{12}\frac{\chi^2}{M^2_P}, 
\ee 
where $m^2_{Y_i}(T)$ is
the thermal mass for the $Y_i$ field which depends on all the
other interactions of $Y_i$.  For instance a coupling to matter
fields $f$ and $\bar f$ in a Yukawa-like fashion $W\supset
\gamma_i Y_i f\bar f$ leads to a thermal mass $ m^2_{Y_i}(T)=
\frac{\gamma_i^2}{6} T^2$.  This is larger than the Hubble rate 
($H \sim \frac{T^2}{M_P}$ in the radiation dominated era) and
drives $Y_i$ to the origin.  As a result, the inflaton $\chi$ gets
an effective mass square, $m^2_{\rm eff} \simeq \sum_{i=1}^P
\sigma_i T^4/({12 M^2_P})$, where for instance $\sigma_i= b_i
\gamma_i^2/6$, which is related to  the Hubble mass squared as 
$m^2_{\rm eff}=p^2 H^2$ in the radiation dominated
pre-inflationary epoch. Therefore solving the evolution equation
for $\chi$, one finds\footnote{Other fields may also have a
thermal mass, but those are irrelevant for our analysis as they
are not destabilising anything.} 
\be \chi = \chi_* \left (
\frac{\rho}{\rho_*}\right )^{1/2} \cos \left ( \sqrt {p^2 - 1/4}
\ln{(\frac{R}{R_*})}\right ), 
\ee 
where $\chi_*$, $R_*$ and
$\rho_*$ represent the amplitude (supposed to be $\sim M_X$ ), the
scale factor and the energy density when the $Y_i$ fields are in
thermal equilibrium at temperature $T_*$. The above expression
implies that once $T < T_*$, the $\chi$ field performs damped
oscillation about $\chi = 0$ and it would continue till the vacuum
energy of inflation is comparable to the radiation density, i.e.
$\rho \sim M^4$. If the amplitude at this point coincides
with the initial value required for $\chi$, $\chi_0 \sim 10^{-4}
M_X$, then the initial condition issue is resolved and inflation
starts. 
Equating
$ \chi_0 = c M_X \left ( \frac{M}{T_*}\right )^{1/2}$  where
$c=  \langle \chi_* \rangle \cos \left ( \sqrt {p^2 - 1/4} \ln{(\frac{R_0}{R_*})}\right ) \lesssim 1$  
includes the ambiguity
of the field value of $\chi_*$ as well as the value of the
oscillatory cosine term,  it follows that with $T_* \sim 10^{18-19}$
GeV  the initial value problem is solved.
Hence for an initial temperature close to the Planck scale, the
initial value of $\chi$ is such that the field starts rolling
slowly and leads inflation at the end of pre-inflationary phase.
Note that the inflaton gets a purely thermal mass through the 
breaking of shift symmetry before inflation, such a term becoming negligible 
during inflation so that the shift symmetry conserving term still 
protects the inflation  mass and solves the $\eta$ problem\footnote{During the
  pre-inflationary era, the $Y_i$ fields are driven to the origin as
  their thermal masses are larger than the Hubble rate. At the end of
  this pre-inflationary epoch and as soon as inflation starts, the
  $Y_i$ fields have a mass term of order $H_I$ which guarantees their
  stability at the origin. As a result, the non-renormalisable term
  $K_+$ does not contribute to the $\eta$ problem despite its
  shift-symmetry breaking feature.}. 

It is worth comparing the present set up with the one in \cite{aru}.
Both are based on $R$-symmetry, and the inflationary superpotential is basically the same,
but thanks to the differences in their K\"ahler potentials,  the inflaton is $S$ in \cite{aru}
and $(\chi + \bar\chi )$ in the present work, and the inflation mechanisms are different.
As a consequence, the inflationary models in \cite{aru} yield a supersymmetry breaking scale
consistent with gravity mediation, while here this scale is much lower and consistent with
gauge mediation. Although we do not tackle the issue in this paper, the needed
couplings of the inflation and supersymmetry breaking to the MSSM fields are expected
to be different as well.
We have studied the gravitational coupling between intermediate (or even low) scale
inflation and ISS metastable supersymmetry breaking in models where 
both phenomena are 
regulated  by an $R$ symmetry. We have found how the
supersymmetry breaking scale is related to the low value of the Hubble rate
during inflation. Requiring the reheating temperature to be 
above the electroweak scale, we obtain bounds on the supersymmetry 
breaking scale as  $10^4$ GeV $\lesssim \mu \lesssim 10^8$ GeV.

\vskip 10pt
    
\noindent {\bf \textit{Acknowledgements}} -  This work is supported by 
the RTN European Program MRTN-CT-2004-503369
and by the French ANR Program PHYS@COL\&COS.


\end{document}